\documentclass[12pt]{iopart}
\usepackage{amssymb}

\begin{document}


\title{Coupling Self-Dual $p$-Form 
  Gauge Fields 
   to Self-Dual Branes}

\author{Chris Hull}

\address{The Blackett Laboratory, Imperial College London, Prince Consort Road, London SW7 2AZ, United Kingdom}
\ead{c.hull@imperial.ac.uk}
\vspace{10pt}
\begin{indented}
\item[]December  2024
\end{indented}

\begin{abstract}
In $d=4k+2 $ dimensions, $p$-form 
  gauge fields (with $p=2k$) with self-dual field strengths couple naturally to dyonic branes with equal electric and magnetic charges.
  Sen's action for a $p$-form gauge field with self-dual field strength coupled to a spacetime metric $g$ involves an explicit Minkowski metric; however, this action can be generalised to provide a theory
  in which the Minkowski metric is replaced by a second metric $\bar g$ on spacetime.  This theory describes  a physical sector, consisting of the chiral $p$-form gauge field coupled to the dynamical metric $g$, plus an auxiliary sector consisting of a second chiral $p$-form and the second metric $\bar g$.
 The fields in this auxiliary sector     only couple to each other and have no interactions with the physical sector.
However, in this theory, the standard coupling to a brane given by integrating the gauge potential over the world-volume of the brane is problematic as the physical gauge potential depends non-locally on the fields appearing in the action. A consistent coupling is given by introducing Dirac branes (generalising Dirac strings), and  is shown to have generalised symmetries corresponding to invariance under deforming the positions of the Dirac branes, provided the Dirac branes do not intersect any physical brane world-volumes.
\end{abstract}

%
%
%
%
%

\section{Introduction}
\label{Introduction}

  Dirac's  quantum theory \cite{Dirac:1948um} of the electromagnetic field in four
dimensions coupling to both electrically charged particles and magnetic monopoles 
requires the introduction of Dirac strings attached to the magnetic monopoles. The positions of the Dirac strings are arbitrary, apart from the requirement that they do 
not intersect the worldlines of any electrically charged particles.
This constraint on the positions of the Dirac strings is sometimes referred to as the Dirac veto. 
In $d$ dimensions, a $p$-form gauge field couples to electrically charged $p - 1$ branes and magnetically charged
$\tilde{p} - 1$ branes with
$\tilde{p} = d - p - 2$ \cite{Nepomechie:1984wu},\cite{Teitelboim:1985ya}.
Dirac's action was generalised to  an action for
$p$-form gauge fields in $d$ dimensions coupling to both electrically and magnetically charged branes by Deser,
Gomberoff, Henneaux and Teitelboim
in \cite{Deser:1997se,Deser:1997mz}. The Dirac strings attached to magnetic monopoles in four dimensions generalise to Dirac $\tilde{p} $ branes
attached to the magnetically charged
$\tilde{p} - 1$ branes, and the Dirac veto now requires that the Dirac branes not intersect the world-volumes of the electrically charged $p - 1$ branes.

In \cite{Hull:2024uwz} it was shown that the theory's independence   of the positions of the Dirac strings or branes could be understood in terms of  generalised symmetries of the theory, with the Dirac veto seen as a restriction to configurations for which a certain anomaly in the generalised symmetries is absent. 

The aim of this paper is to extend this analysis to the theory of self-dual $p$-form gauge fields coupled to self-dual branes.
This requires $p$ to be even, $p=2k$, and the dimension to be $d=4k+2$, so that $\tilde{p} = p$ (with Lorentzian signature).
For the self-dual theory, the $p+1$-form field strength $F$ is self-dual, $F=*F$ and this couples to dyonic  $p$-branes with equal electric and magnetic charges.
For  the 4-form gauge field in IIB supergravity  (with $p=4$) the  coupling would be to a D3 brane in 10 dimensions while for $p=2$  a 2-form gauge field  would couple to a self-dual string in 6 dimensions. For $p=0$  the theory gives a right-moving scalar in 2 dimensions.

The construction of an action for an antisymmetric tensor gauge field with
self-dual field strength is a problem that has attracted a great deal of
attention, and many approaches have been used; see  e.g.\ \cite{Evnin:2022kqn},\cite{Hull:2023dgp}   for a list of references; for a recent review and critical comparison of the main approaches,  see \cite{Evnin:2022kqn}. 
An approach that has attracted a lot of attention is the PST action \cite{Pasti:1996vs} and the coupling of this to self-dual  branes, using  Dirac's formulation,  was given
in \cite{Medina:1997fn},
\cite{Lechner:1999ga},
\cite{Lechner:2000eg}.

In  \cite{Sen:2015nph,Sen:2019qit}, Sen constructed an interesting action for self-dual antisymmetric tensor 
 gauge fields, which was 
inspired by the string field theory for the IIB superstring. This  approach is covariant and the action is quadratic in the fields, facilitating quantum calculations, and it  also generalises to allow interactions; it is further discussed in 
\cite{Evnin:2022kqn,Andriolo:2020ykk,Vanichchapongjaroen:2020wza, Andrianopoli:2022bzr,Chakrabarti:2022jcb,Barbagallo:2022kbt,Chakrabarti:2020dhv,Andriolo:2021gen}. 
Sen's action gives a self-dual $p$-form gauge  field coupling to the space-time metric $g$ and other physical fields, together with a second self-dual $p$-form gauge  field which doesn't couple to the space-time metric or any  physical fields, but which instead couples to a Minkowski metric.

Sen's action was generalised in \cite{Hull:2023dgp} to an action in which the second self-dual $p$-form gauge  field couples to an arbitrary second metric $\bar g$ 
instead of the Minkowski metric. This means that   the action can be formulated on any spacetime (not just spacetimes admitting a Minkowski metric) and gives  a theory with two gauge invariances corresponding to the two gauge fields $g,\bar g$.  

In this article, the action for self-dual gauge fields of \cite{Hull:2023dgp} will be coupled   to self-dual branes and the resulting  generalised symmetries will be investigated.  
However, this coupling is not straightforward, as will now be discussed.
A $p$-form gauge field $A$ typically couples to $p-1$ branes with an electric coupling of the form
\begin{equation}
\label{brane}
S_{brane} =\mu \int _\mathcal{N} A
\end{equation}
where $\mathcal{N}$ is the $p$ dimensional submanifold on which the
brane is located and $\mu$ is the charge of the  brane.
If the brane also carries magnetic charge, then there must also be a magnetic coupling to the brane, which can be formulated using the Dirac approach \cite{Dirac:1948um} that will be reviewed in the next section.
If the gauge field $A$ has self-dual field strength, then the action of \cite{Hull:2023dgp} can be used for the free theory, but then
 the coupling $\mu \int A$ is problematic as $A$ is not a fundamental field in the action and, as will be seen in section \ref{Action}, it is the field strength $F$ that has a local expression in terms of the fundamental fields.
  However, the coupling (\ref{brane}) can be rewritten as
  \begin{equation}
\label{branes}
S_{brane} =\mu \int _\mathcal{P} F
\end{equation}
  where $F=dA$ and  $\mathcal{P}$ is a $p$-dimensional subspace with boundary $\mathcal{N}$.
  The functional integral is then independent of the choice of $\mathcal{P}$ provided the flux of $\mu F$ through any closed $p+1$ surface is quantised  appropriately, as is the case in string theory. Then $F$ can be rewritten in terms of the fundamental fields in the action, so that (\ref{branes}) gives a coupling of the self-dual gauge field to the brane.
  Such a coupling will be derived in   sections \ref{Action},\ref{Coupling} and shown to give the desired field equations. An alternative coupling of Sen's action to branes was 
  given in \cite{Lambert:2023qgs}; this coupling required that the flux of the non-physical gauge field vanish.
  
  This coupling applies to two situations.
  In the first, $\mathcal{N}$ is a cycle that is the boundary of some 
$p+1$-dimensional submanifold $\mathcal{P}$, $\mathcal{N}=\partial \mathcal{P}$.
For $p=1$ this gives a Wilson line on a closed curve $\mathcal{N}$ bounding a disk
$\mathcal{P}$, while for $p>1$ this gives a Wilson $p$-surface.
In the second situation, for $p=1$ $\mathcal{N}$ is the world-line of a particle  
and $\mathcal{P}$ is the world-sheet of a Dirac string from the particle to infinity, while for $p>1$  $\mathcal{N}$ 
is the  world-volume of a $p-1$ brane and 
  $\mathcal{P}$ is the world-volume of a Dirac $p$ brane ending on the $p-1$ brane.
  In both situations, it will be useful to refer to $\mathcal{P}$ as the location of a Dirac brane.

\section{Antisymmetric Tensor Gauge Fields with Sources}
\label{Antisymmetric}

A $(p +
1)$-form field strength $F$ in a $d$ dimensional   spacetime satisfies the
equations
\begin{equation}
  dF = \ast \tilde{j}  \qquad d \ast F = \ast j \label{Ffidaa}
\end{equation}
where $j$ is a  $p$-form electric current and $\tilde{j}$ is a
  $\tilde{p}$-form magnetic current, with
$\tilde{p} = d - p - 2$. Both currents are required to be conserved,
\begin{equation}
d^\dag j=0, \qquad d^\dag \tilde j=0\, .
\end{equation}

In the absence of magnetic sources, i.e.\ with $\tilde{j}=0$,  $dF=0$ so that locally there is a $p$-form gauge potential $A$ with $F=d A$. The action is then
\begin{equation}
S=\int F\wedge *F - A\wedge *j
\label{act}
\end{equation}
In general, if $A$ is defined locally, with  different potentials $A$ in different coordinate patches, then $\int   A\wedge *j$ is ill-defined, and a definition such as that in  \cite{Wu:1976qk},\cite{Alvarez:1984es} is used instead, giving rise to the Dirac quantisation condition.
If the magnetic current $\tilde j$ is non-zero but the electric current $j$ vanishes, then 
there is a similar treatment as a theory of a magnetic potential $\tilde A$ with $F=\ast d \tilde A$.

Consider now the general case with both electric and magnetic sources, with both 
 $j$ and   $\tilde j$   non-zero.
Following the approach of Dirac \cite{Dirac:1948um} and Deser {\it {et al}}, \cite{Deser:1997se,Deser:1997mz},   the equation $dF = \ast \tilde{j}$ can be solved by introducing a $(\tilde p+1)$-form current
$\tilde{J}$ satisfying
\begin{equation}
  \begin{array}{lll}
    {d }^{\dag} \tilde{J} & = & \tilde{j}
  \end{array} \label{Jjt}
\end{equation}
so that
\begin{equation}
  d (F - \ast \tilde{J}) = 0
\end{equation}
and there is a $p$-form potential $A$ satisfying
\begin{equation}
  F = \ast \tilde{J} + d A \label{fja}
\end{equation}
Note that this requires that the current $\tilde j$ is conserved off-shell, which is the case for magnetically charged branes.

For Maxwell theory in $d=4$ with $p=1$, 
the $1$-form $\tilde{j}$ is the magnetic monopole current. A Dirac string is
attached to each magnetic monopole and the $2$-form $\tilde{J}$ is the current
density for these strings: if  $\tilde{j}$ is localised on the world-line of
a magnetic monopole, then $\tilde{J}$ is localised on the world-sheet of the
corresponding Dirac string. For general $d,p$, if the $\tilde p$-form current $\tilde{j}$ is the magnetic brane current localised on the world-volume of
a magnetic $\tilde p -1$ brane, then the $(\tilde p+1 )$-form current $\tilde{J}$  is the Dirac brane current localised on the $(\tilde p+1 )$-dimensional world-volume of
a Dirac $\tilde p $-brane ending on the magnetic $\tilde p -1$ brane.

  Dirac's action is given by the sum of the kinetic terms for the electric
and magnetically charged particles plus  (\ref{act})
with $F = \ast \tilde{J} + d A$. This gives the correct field equations,
provided that the condition that has become known as the
{\emph{Dirac veto}} holds. This requires that the
positions of the Dirac $\tilde p $-branes be restricted so that there is no intersection
between the world-volumes of the electric  $p-1$-branes and the world-volumes
of the Dirac $\tilde p $-branes. In particular, the field equations do not depend on the
locations of the Dirac branes provided that they comply with the Dirac veto,
and so do not depend on the choice of $\tilde{J}$ satisfying (\ref{Jjt}).
If there are no magnetic sources, then $\tilde{j}=0$ and $\tilde{J}=0$ and the theory reduces to the usual Maxwell action.

Dirac's action is not single-valued. A continuous deformation of the
positions of the Dirac branes (while obeying the veto) can change the action by any integral multiple
of $4 \pi q p$ where $q$ is the electric charge of any electric brane and $p$ is the
magnetic charge of any magnetic brane \cite{Dirac:1948um,  Deser:1997se,Deser:1997mz}. Then $e^{i S / \hbar}$ will be single
valued provided the electric and magnetic charges all satisfy the Dirac
quantisation condition and the quantum theory is then well-defined.

Dirac branes can instead be introduced  for the
electrically charged branes. If the electric current is $j$, there is then
a $( p+1)$-form current $J$ localised on the world-volumes of the electric Dirac
branes satisfying
\begin{equation}
  {d }^{\dag} J = j
   \label{djis}
\end{equation}
Then
\[ d^{\dag} F = j \]
can be written as
\begin{equation}
  d (\tilde{F} - \ast J) = 0
\end{equation}
(writing $\tilde{F} = \ast F$ for the Hodge dual of $F$) so that there is a dual
formulation in terms of a dual potential $\tilde{A}$ with
\begin{equation}
  \tilde{F} = \ast J + d \tilde{A}
\end{equation}
with action
\begin{equation}
  S [\tilde{A}] = \int \frac{1}{2} \tilde{F} \wedge \ast \tilde{F} - \tilde{A}
  \wedge \ast \tilde{j} \label{duact}
\end{equation}
In this case, Dirac's veto requires that the electric Dirac branes on which
the current $J$ is localised do not intersect the world-volumes of the
magnetically charged branes; this will be referred to as the dual Dirac veto.

The secondary current
$J$ satisfying
(\ref{djis})
 can be used to rewrite the electric coupling  as
\begin{equation}
 \int   A\wedge *j=-\int   F\wedge *J+ \int   \tilde J \wedge *J
\end{equation}
The term $\int   \tilde J \wedge *J$ is independent of $A$ and depends only on the matter fields and can be absorbed into the action for these, leaving the action
\begin{equation}
  \hat{S} [A] = \int \frac{1}{2} F \wedge \ast F + F \wedge \ast J
  \label{acth}
\end{equation}

If $A$ is only locally-defined,  $\int   F\wedge *J$ will be well-defined if $J$ is well-defined and gives a covariant coupling. However, 
  for a given $j$, different choices of $J$ satisfying (\ref{djis})    give different actions in general, giving rise to an ambiguity.
For the case in which the current is carried by charged branes, requiring the path integral be unambiguous gives rise to the Dirac quantisation condition \cite{Hull:2024uwz}.

\section{Charged Branes}
\label{Charged}

If the source is  an  electrically charged $p - 1$ brane
whose  world-volume is a $p$-dimensional submanifold $\mathcal{N} \subset
\mathcal{M}$, then the coupling can be written as
\begin{equation}
  q \int_{\mathcal{N}} A = \int_{\mathcal{M}} A \wedge \ast j \label{qA2}
\end{equation}
The current is localised on  $\mathcal{N}$ and can be written as
\begin{equation}
  j = q \delta_{\mathcal{N}}
\end{equation}
where $q$ is the electric charge of the brane and  $\delta_{\mathcal{N}}$ can be viewed as a $p$-form with components given by  delta-functions so that (\ref{qA2}) holds.
Singular forms such as
$\delta_{\mathcal{N}}$ are examples of what mathematicians call
{\emph{currents}}, as defined in \cite{Rham,Griffiths}.\footnote{
Some of the formulae in this paper involve products of currents. If these are
delta-function currents, such products can be ill-defined. As in \cite{Dirac:1948um,Deser:1997se,Deser:1997mz,Hull:2024uwz}  it
will be supposed here that the delta functions are smeared to some smooth functions
where necessary.}
If $\mathcal{N} \subset \mathcal{M}$ is specified by $x^{\mu} = X^{\mu}
(\sigma^a)$ for some functions $X^{\mu}
(\sigma^a)$ of the
world-volume coordinates
 $\sigma^a$ ($a = 0, 1, \ldots, p - 1$),
 then the current has components
\begin{equation}
  j^{\mu_1 \ldots \mu_{q - 1}} = q  \int d^{q - 1} \sigma  \quad
  \varepsilon^{a_1 a_2 \ldots a_{q - 1}}  \frac{\partial X^{\mu_1}}{\partial
  \sigma ^{a_1}} \frac{\partial X^{\mu_2}}{\partial \sigma ^{a_2}} \ldots
  \frac{\partial X^{\mu_{q - 1}}}{\partial \sigma ^{a_{q - 1}}} \delta (x - X
  (\sigma )) \, .\label{jcuris}
\end{equation}

Consider first the case in which
  $\mathcal{N}$  is a cycle that is the boundary of some 
$p+1$-dimensional submanifold $\mathcal{P}$, $\mathcal{N}=\partial \mathcal{P}$, then 
the coupling (\ref{qA2}) can be rewritten as
\begin{equation}
  q \int_{\mathcal{N}} A = q \int_{\mathcal{P}} F \label{afcd}
\end{equation}
which can be re-expressed as
\begin{equation}
  \int_{\mathcal{M}} A \wedge \ast j = \int_{\mathcal{M}} F \wedge \ast J
  \label{ajfjis}
\end{equation}
where
\begin{equation}
  J = q \delta_{\mathcal{P}}
\end{equation}
and satisfies $d^{\dag} J = j$ as a result of
\begin{equation}
  \delta_{\partial \mathcal{P}} = d^{\dag} \delta_{\mathcal{P}}
  \label{ddagdel}
\end{equation}
This then gives a construction of the secondary current $J$ satisfying (\ref{djis}).

However, $q
\int_{\mathcal{P}} F$ depends on the choice of surface with boundary
$\mathcal{N}$. For two surfaces $\mathcal{P}, \mathcal{P}'$ with boundary
$\mathcal{N}$,
\begin{equation}
  q \int_{\mathcal{P}'} F - q \int_{\mathcal{P}} F = q \int_{\mathcal{Q}} F
\end{equation}
where $\mathcal{Q}=\mathcal{P}  \cup \mathcal{P}'$ is the closed surface given
by combining $\mathcal{P}, \mathcal{P}'$ with opposite orientations. Note that
$p = \int_{\mathcal{Q}} F$ is the magnetic charge contained in $\mathcal{Q}$
(which is $2 \pi$ times an integer if $F$ is conventionally normalised). Then
the Wilson surface
\begin{equation}
  W (\mathcal{N}) = e^{\frac{i}{\hbar} q \int_{\mathcal{P}} F}
\end{equation}
changes by a phase
\[ e^{\frac{i}{\hbar} q p} \]
on changing from $\mathcal{P}$ to $\mathcal{P}'$ and so is well-defined
provided that the charges satisfy the Dirac quantisation condition
\begin{equation}
  p q = 2 \pi n, \quad n \in \mathbb{Z}
\end{equation}
for some integer $n$.
See \cite{Hull:2023dgp} for further discussion.

Now consider the case in which $\mathcal{N}$ is not a cycle but is the
world-volume of a physical charged brane. For example, for a   charged particle, $\mathcal{N}$ is the particle world-line $X^\mu(\tau)$.  (The charge could be electric or magnetic.) For each $\tau$,    a Dirac string is introduced that
goes from the particle to infinity and which is specified by functions $Y^{\mu} (\tau,
\sigma)$ with $Y^{\mu} (\tau, 0) = X^{\mu} (\tau)$ so that the string
world-sheet $\mathcal{P}$ is specified by $x^{\mu} = Y^{\mu} (\tau, \sigma)$.
The boundary of $\mathcal{P}$ is $\mathcal{N} \cup \mathcal{G}$ where
\ensuremath{\mathcal{G}} is the part of the boundary at infinity.

For such a world-line, (\ref{afcd}) becomes
\begin{equation}
  q \int_{\mathcal{N}} A = q \int_{\mathcal{P}} F - q \int_{\mathcal{G}} A
\end{equation}
and (\ref{afcd}) only holds with suitable boundary conditions, e.g.\ if $A = 0$
on $\mathcal{G}$, in which case one has (\ref{ajfjis}). Note that the actions
$q \int_{\mathcal{N}} A$ and $q \int_{\mathcal{P}} F$ give the same field
equations from variations that vanish on $\mathcal{G}$. This then generalises to the case of general $p,d$ with $\mathcal{N}$ the world-volume of a magnetically charged brane and $\mathcal{P}$ the world-volume of a Dirac brane ending on $\mathcal{N}$.
For a magnetically charged brane, the coupling of the dual potential $\tilde A$ is
\begin{equation}
  p \int_{\mathcal{N}} \tilde A = q \int_{\mathcal{P}} \ast F \label{afcda}
\end{equation}
which can be re-expressed as
\begin{equation}
  \int_{\mathcal{M}} \tilde A \wedge \ast \tilde j = \int_{\mathcal{M}} F \wedge  \tilde J
  \label{ajfjisa}
\end{equation}

\section{Generalised Symmetries}
\label{Generalised}

The actions of Dirac and Deser {\it {et al}} reviewed in  section \ref{Antisymmetric}
give field equations that do not depend on the position of the
Dirac strings or branes, provided that  they comply with the Dirac veto. 
In
 \cite{Hull:2024uwz},
 the dependence of the action on the position of the strings was investigated and it
 was shown that the action  is invariant under changing the
positions of the Dirac strings (subject to the Dirac veto)
 and that this invariance can be formulated  in terms of  extra gauge symmetries of the action.
These are $p$-form and $\tilde p$-form
generalised symmetries and   the Dirac veto arises as a
condition for the absence of anomalies in these generalised   symmetries.

The equations
\begin{equation}
  {d }^{\dag} J = j,  \quad \begin{array}{lll}
    {d }^{\dag} \tilde{J} & = & \tilde{j}
  \end{array}
\end{equation}
don't determine the currents $J, \tilde{J}$ uniquely: they can be transformed by
\begin{equation}
  \delta J = d^{\dag} \rho, \quad \delta \tilde{J} = d^{\dag} \tilde{\rho}
  \label{Jro}
\end{equation}
for some $(p+2)$-form $\rho$ and $(\tilde p+2)$-form $ \tilde{\rho}$. 
In order for $F = \ast \tilde{J} + d A$ and $\tilde{F} = \ast J + d
\tilde{A}$ to remain invariant, it is then necessary that the potentials shift
under these transformations as
\begin{equation}
  \delta A = \ast \tilde{\rho}, \quad \delta \tilde{A} = \ast \rho
   \label{arho}
\end{equation}
Dualising  $\rho = \ast \tilde{\lambda}$, $\tilde{\rho} = \ast \lambda$ gives a
$p$-form parameter $\lambda$ and a $\tilde p$-form parameter $ \tilde{\lambda}$, so that the transformations become
\begin{eqnarray}
  \delta A = \lambda, &  & \delta \tilde{J} = \ast d \lambda  \label{lam}\\
  \delta \tilde{A} = \tilde{\lambda}, &  & \delta J = \ast d \tilde{\lambda} 
  \label{lamt}
\end{eqnarray}
Note that each of the actions that have been discussed depend only on $A$ or only on $\tilde A$ but not both.

The interpretation of these transformations  is
as follows for a magnetic Dirac brane. Smoothly deforming the $\tilde p$-dimensional submanifold $\mathcal{P} $ on which a Dirac
brane is localised to a  submanifold $\mathcal{P} '$ gives a family of Dirac
brane world-volumes $\mathcal{P} (\xi)$ parameterised by $\xi\in [0,1]$ with
$\mathcal{P} (0) =\mathcal{P}$ and $\mathcal{P} (1) =\mathcal{P}'$. This
family of   world-volumes sweeps out a $(\tilde p+1)$-dimensional   submanifold
$\mathcal{Q}$. For a magnetic Dirac brane, the resulting change in $\tilde{J}
$ is, for an infinitesimal deformation of $\mathcal{P}$, of the form $\delta
\tilde{J} = d^{\dag} \tilde{\rho}$ where $\tilde{\rho}$ is a   current 
localised on $\mathcal{Q}$.  The position of each magnetic Dirac brane $\mathcal{P} (\xi)$
should satisfy the Dirac veto, so that each $\mathcal{P} (\xi)$
 should not intersect the world-volume of any
electric brane, and so $\mathcal{Q}$ should not intersect the world-volume of any
electric brane. As $\tilde{\rho}$ is a   current 
localised on $\mathcal{Q}$ and $j$ is localised on the electric brane world-volumes, the Dirac veto implies
\begin{equation}
\int j \wedge \tilde{\rho} =0
\label{veto}
\end{equation}

The situation is similar for a deformation of an electric Dirac brane (with $\tilde p$ replaced by $p$): the change in
$J$ is $\delta J = d^{\dag} \rho$ where $\rho$ is a $(p+1)$-form current localised
on the $(  p+1)$-dimensional   submanifold
 swept out by the family of Dirac branes.

The variation of the action (\ref{act}) under (\ref{lam}) is
\begin{equation}
  \delta S = - \int \lambda \wedge \ast j \label{varl}
\end{equation}
which, using using $\tilde{\rho} = \ast \lambda$,  vanishes as a result of the Dirac veto condition (\ref{veto}). In other words, 
if the theory is restricted to configurations consistent with the Dirac veto,
then none of the family of Dirac branes $\mathcal{P} (\xi)$ intersect the
world-volumes of electric branes and this implies that $\tilde{\rho} = \ast \lambda$ is restricted to vanish at
any place where $j$ is non-zero. As a result, the Dirac veto condition (\ref{veto}) ensures that the
variation (\ref{varl}) vanishes and the action is invariant under
(\ref{lamt}).

The alternative action (\ref{acth}) is invariant under (\ref{lam}) but under
(\ref{lamt}) it transforms as
\begin{equation}
  \delta \hat S = \int F \wedge d \tilde{\lambda}
\end{equation}
Here $F$ is {\emph{defined}} by (\ref{fja}) and so
\begin{equation}
  d F = \ast \tilde{j}
\end{equation}
and as a result
\begin{equation}
  \delta \hat S = \int \tilde{\lambda} \wedge \ast \tilde{j}
\end{equation}
 This now vanishes as a result of the dual Dirac veto, using $\rho = \ast \tilde{\lambda}$. Similarly, the dual action
(\ref{duact}) is invariant under (\ref{lam}),(\ref{lamt}) provided that
the   Dirac veto for the electric Dirac branes holds.

Then the   action has generalised symmetries corresponding to the symmetry
under changing the
positions of the Dirac branes.
Remarkably, the structure outlined above also appears in the study of
generalised symmetries of Maxwell theory  and in its extension to $p$-form gauge fields in $d$ dimensions \cite{Gaiotto:2014kfa}; see e.g.\cite{Bhardwaj:2023kri,Brennan:2023mmt,Schafer-Nameki:2023jdn} for reviews and an extensive list of references.
For example, in $d=4$, Maxwell theory (without sources) has a
$1$-form symmetry $\delta A = \lambda$ with $d \lambda = 0$. This can be
gauged, i.e.\ promoted to a symmetry for general $\lambda$, by coupling to a
$2$-form gauge field $B$, so that the gauge-invariant field strength is $F = d
A - B$. This agrees with (\ref{fja}) if one takes $B = - \ast \tilde{J}$, so
that the Dirac string current can be interpreted as (the dual of) a   gauge field.
There is a similar story for gauging the dual $1$-form symmetry $\delta
\tilde{A} = \tilde{\lambda}$ with gauge field $\tilde{B}$, which can be
identified with $- \ast J$. There is an obstruction to gauging both of these
$1$-form symmetries simultaneously, and this is often expressed by saying
that these symmetries have a mixed anomaly. Then the Dirac veto can be viewed
as a restriction to configurations of the   gauge fields $B,
\tilde{B}$ for which the anomaly vanishes.

\section{The Self-Dual Theory}
\label{Self}

The analysis of the previous sections will now be applied to the case in which 
$d=4k+2$ and $p=\tilde p=2k$ with self-dual sources, i.e.\  $j=\tilde j$ and so 
$J=\tilde J$. Then (\ref{fja}) becomes
\begin{equation}
  F = \ast  {J} + d A \label{fja2}
\end{equation}
which satisfies
\begin{equation}
 dF = \ast  {j} 
 \label{dfja}
\end{equation}
and the action is (\ref{act}) or (\ref{acth}) with field equation
\begin{equation}
 d\ast F = \ast  {j} 
\end{equation}
For the self-dual theory, these equations are supplemented by the constraint
\begin{equation}
F=\ast F
\end{equation}
which is consistent with the above equations.

Consider a set of  $N$ self-dual $p-1$ branes with charges $q_i$ located on
$p$-submanifolds ${\mathcal{N}_i}$. As before, two cases will be considered. In the first, each ${\mathcal{N}_i}$ is a cycle that is the boundary of a  ${\mathcal{P}_i}$, $\partial {\mathcal{P}_i}={\mathcal{N}_i}$. In the second, there is a Dirac brane with $p+1$ dimensional world-volume ${\mathcal{P}_i}$ attached to each brane.

The current is then
\begin{equation}
  j (x) = \sum_{i = 1}^N q_i \delta_{\mathcal{N}_i} (x),  \label{jiss}
\end{equation}
which satisfies 
\begin{equation}
   {j} = d^{\dag}  {J}
  \label{wetws}
\end{equation}
where the secondary current is
\begin{equation}
   {J} = \sum_i q_i \delta_{\mathcal{P}_i}
  \label{Jtiis}
\end{equation}
and is localised on the Dirac branes at ${\mathcal{P}_i}$.
The Dirac veto for this case was considered in \cite{Deser:1997se},\cite{Deser:1997mz},\cite{Hull:2024uwz} and restricts the location of the  Dirac brane ${\mathcal{P}_i}$ to not intersect the world-line of any other brane world-volume ${\mathcal{N}_j}$:
\begin{equation}
  \mathcal{P}_i \cap \mathcal{N}_j = 0 \qquad  {\rm{for}} \quad j \neq i
  \label{defvet}
\end{equation}
If the $i$'th Dirac brane is deformed as before to sweep out a surface ${\mathcal{Q}_i}$, these are required to satisfy
\begin{equation}
  \mathcal{Q}_i \cap \mathcal{N}_j = 0 \qquad  {\rm{for}} \quad j \neq i
  \label{defveta}
\end{equation}
Defining
\begin{equation}
   {\rho} = \sum_i q_i \delta_{\mathcal{Q}_i}
  \label{rhoiis}
\end{equation}
the constraint (\ref{defveta}) can be written as \cite{Hull:2024uwz}
\begin{equation}
\label{rhveto}
\int j\wedge \rho =0
\end{equation}
Here, the terms involving $\delta_{\mathcal{Q}_i}\wedge \delta_{\mathcal{N}_j}$
for $i=j$ were shown to vanish in \cite{Hull:2024uwz} provided the delta-functions are suitably regularised.

The field strength is invariant under the 
transformations
\begin{eqnarray}
  \delta A = \lambda, &  &  \delta J = \ast d  {\lambda} 
  \label{lamta}
\end{eqnarray}
Under these transformations, the action (\ref{acth}) transforms by
\begin{equation}
\label{erhdgh}
\delta \hat S=\int F\wedge d  {\lambda} =- \int j\wedge  \ast {\lambda}
\end{equation}
using (\ref{dfja}). Then this vanishes for variations with ${\lambda}=\ast \rho$ with $\rho$ of the form 
(\ref{rhoiis}) provided that the Dirac veto constraint (\ref{rhveto}) holds.

In this section, the self-duality condition $F=\ast F$ was introduced as an additional constraint
 that is consistent with the field equations. In the following sections, the analysis will be revisited using an action that gives the self-duality condition  as a field equation.

\section{Action for gauge fields with Sources}
\label{Action}

 Sen's action for a $p$-form gauge field with self-dual field strength coupled to a spacetime metric $g$ involves an explicit Minkowski metric and the presence of this raises questions as to whether the action is coordinate independent and whether it can be used on a general spacetime manifold.
A   generalisation of Sen's action was presented in \cite{Hull:2023dgp} in which the Minkowski metric is replaced by a second metric $\bar g$ on spacetime. The theory is  covariant and can be formulated on any spacetime.
 The theory describes  a physical sector, consisting of the chiral $p$-form gauge field  $A$ coupled to the dynamical metric $g$ and any other physical fields, plus a shadow sector consisting of a second chiral $p$-form $C$ and the second metric $\bar g$.
 The fields in this shadow sector     only couple to each other and have no interactions with the physical sector, so that they decouple from the physical sector.
 
 In addition to the Hodge dual $\ast$ with respect to the spacetime metric $g$, there is a second Hodge dual  $\bar{\ast}$ 
with respect to the second metric $\bar g$. The physical field strength $F=dA+\dots$ is self-dual with respect to the spacetime metric $g$, $F=\ast F$, while the shadow-sector field strength $G=dC$ is self-dual with respect to the other metric $\bar g$, $G = \bar{\ast} G$.
The action has two diffeomorphism-like symmetries, one acting only on the physical sector 
and one acting only on the shadow sector, with the spacetime  diffeomorphism symmetry arising as the diagonal subgroup. It will be useful to introduce projectors acting on $p+1$-forms in dimension $d=2p+2$: 
\begin{equation}
  \bar{\Pi}_{\pm} = \frac{1}{2}  (1 \pm \bar{\ast}), \quad \Pi_{\pm} =
  \frac{1}{2}  (1 \pm \ast)
\end{equation}

The physical gauge field will be taken to couple to matter fields  through a $p+1$-form $\Omega$,
resulting in a
 self-dual field strength $$F = d A + \Omega$$
 The action is written in terms of
   a $p$-form
$P$ and a $p+1$-form $Q$ that  
 is self-dual with respect to the   metric $\bar g$,
 $Q =\bar \ast Q$. The field strengths $F,G$ are then constructed from the dynamical fields $P$ and $Q$, as will be seen below.
 The action with coupling to $\Omega$ 
 is \cite{Hull:2023dgp}
\begin{equation}
S=S_0+S_\Omega +S_m
\label{Stot}
\end{equation}
where 
\begin{equation}
S_0= \int \left( 
\frac{1}{2}  d P \wedge \bar{\ast} d P - 2 Q
  \wedge d P - Q \wedge M (Q) \right)
\end{equation}
\begin{equation}
S_\Omega= \int \left( 
 2Q \wedge  \Omega _-
   -2 \Omega_+
  \wedge M (Q ) \right)
  \label{Som}
\end{equation}
and $S_m$ is the action for any other matter fields and the dynamical graviton $g$; $S_m$    depends on the metric $g$ but is independent of $Q,P,\bar g$. 
Here \[ \Omega_{\pm} = \bar{\Pi}_{\pm} \Omega \]
and $M$ is a
linear map on $q=p+1$-forms $Q$ which can be written in components as
\begin{equation}
  \label{Mcomps} M (Q)_{\mu_1 \ldots \mu_q} = \frac{1}{q !} M_{\mu_1
  \ldots \mu_q}^{\nu_1 {\ldots \nu_q} } Q_{\nu_1 \ldots \nu_q}
\end{equation}
for some coefficients $M_{\mu_1 \ldots \mu_q}^{\nu_1 {\ldots \nu_q} } (x)$.
The  coefficients $M_{\mu_1 \ldots \mu_q}^{\nu_1 {\ldots \nu_q} } (x)$ depend on the metrics $g,\bar g$ and are given in \cite{Hull:2023dgp}.
Note that the metric $g$ only enters the actions $S_0,S_\Omega$  through $M (Q)$.  
  The action (\ref{Stot}) reduces to Sen's action  \cite{Sen:2015nph,Sen:2019qit}
for $\bar{g} = \eta$.\footnote{The action (\ref{Stot}) agrees with the action (10.18) of \cite{Hull:2023dgp} (with the parameter $\lambda$ in \cite{Hull:2023dgp} set to zero) up to $\Omega^2$ terms that are independent of the gauge fields and can be absorbed into the action for the matter fields.}

It was shown in \cite{Hull:2024uwz}, using arguments in \cite{Andriolo:2020ykk}, that the map $M$ has the following properties.
 $M$  is symmetric in the sense that
\begin{equation}
  \label{symm} R \wedge M (Q) = Q \wedge M (R)
\end{equation}
for any two \ $p+1$-forms $Q, R$ which are $\bar g$-self-dual, $Q = \bar{\ast} Q$, $R =
\bar{\ast} R$. Moreover, $M(Q)$  is then  $\bar g$-anti-self-dual,
\begin{equation}
  \label{mqsd} \bar{\ast} M (Q) = - M (Q)
\end{equation}
The map $M$ is important as it gives a map from a form $R$ that is self-dual with respect to $\bar g$, $R =
\bar{\ast} R$, to a form that that is self-dual with respect to $  g$,
\begin{equation}
{\Pi}_+R=R+M(R)
\label{pir}
\end{equation}

The field equations for $P, Q$ (using the symmetry and linearity of $M$) are
\begin{equation}
  \label{pfom} d \left( \frac{1}{2}  \bar{\ast} d P + Q + \lambda \Omega
  \right) = 0
\end{equation}
and
\begin{equation}
  \label{qfom} 
  \frac 1 2 (d P-\bar{\ast} d P )+ M (Q+\Omega_+)-\Omega_-=0
\end{equation}
The field strength   defined by
\begin{equation}
  G \equiv \frac{1}{2}  (d P + \bar{\ast} d P) + Q
\end{equation}
  is self-dual
\begin{equation}
  \bar{\ast} G = G
  \label{fcon1}
\end{equation}
and, from (\ref{pfom}), is closed
\begin{equation}
  dG = 0
  \label{fcon2}
  \end{equation}
  so that locally there is a $p$-form potential $C$ with 
  \begin{equation}
G=dC
\label{GdC}
\end{equation}

Taking the exterior derivative of (\ref{qfom}) and eliminating $P$ using
(\ref{pfom}) gives
\begin{equation}
  d [Q + M (Q + \Omega_+)] = d \Omega_-  
  \label{dqm}
\end{equation}
Let
\begin{equation}
  F \equiv Q + \Omega_+ + M (Q + \Omega_+)
\end{equation}
so that, from (\ref{pir}),
\begin{equation}
  F = \Pi_+ (Q + \Omega_+) \, .
  \label{fpi}
\end{equation}
Then from (\ref{dqm})
\begin{equation}
  d F =  d \Omega
  \label{fcon3}
\end{equation}
and, from (\ref{fpi}), $F$ is \ $g$-self-dual,
\begin{equation}
  \ast F = F \, .
  \label{fcon4}
\end{equation}
Then a potential $A $ can be introduced so that
\begin{equation}
  F = d A +   \Omega 
  \label{FdA}
\end{equation}
Then $G=dC$ is a free field coupling only to $\bar g$ so that the shadow sector can be taken to be $\bar g, C$. The physical gauge field $A$ then couples to other physical fields through $F = d A + \Omega$.

The transformations
\begin{equation}
  \label{symom22} \delta P = \lambda, \quad \delta \Omega = d \lambda, \quad
  \delta Q = - \bar{\Pi}_+ d \lambda
\end{equation}
leave the field strengths   invariant, $\delta F = \delta G = 0,$
but the variation of the action under these  is
\begin{equation}
  \label{delss2s} \delta S = -  \int \lambda \wedge d \Omega
\end{equation} 
The field equations are invariant under (\ref{symom22}) but the action
is invariant only under transformations for which (\ref{delss2s}) vanishes.

\section{Coupling to Branes and Generalised Symmetries}
\label{Coupling}

The integrand in  $S_\Omega$ (given in (\ref{Som})) can be rewritten using (\ref{symm}),(\ref{mqsd}):
$$
   Q \wedge  \Omega _-
   - \Omega_+
  \wedge M (Q ) 
  =Q\wedge(\Omega-M(\Omega)
  )=
  - \Omega \wedge (Q+M (Q ))=\Pi_+Q\wedge\Pi_-\Omega
  $$
  so that 
\begin{equation}
S_\Omega= -2\int  
 \Omega \wedge (Q+M (Q ))
\end{equation}
Using (\ref{fpi}),(\ref{FdA}), this differs from the action
\begin{equation}
\label{sdfsr}
S'_\Omega= -2\int  
 \Omega \wedge F
\end{equation}
by a term 
$$-2\int  
\Pi_- \Omega \wedge \Pi _+\Omega  =-2\int  
 \Omega \wedge\ast\Omega
$$
 which does not contribute to the field equations for $P$ or $Q$ and can be absorbed into the matter action $S_m$. Then using $S'_\Omega$
instead of $S_\Omega$ results in the same analysis as in the last section, again leading to field strengths $F,G$ given by (\ref{FdA}),(\ref{GdC}) and satisfying (\ref{fcon1}),(\ref{fcon2}),(\ref{fcon3}),(\ref{fcon4}).

This can now be used to give an action for a field strength $F$ satisfying
\begin{equation}
F=\ast F, \qquad dF=\ast j
\end{equation}
with 
\begin{equation}
j=d^\dag J
\end{equation}
by setting $\Omega =\ast J$ in the above.
Then for a brane of charge $q$ with world-volume   a   submanifold ${\mathcal{N}}$, the current is
\begin{equation}
  j = q \delta_{\mathcal{N}}
\end{equation}
and
\begin{equation}
  J = q \delta_{\mathcal{P}}
\end{equation}
for some submanifold $\mathcal{P}$ with boundary  
$\mathcal{N}=\partial \mathcal{P}$.

The action is then (\ref{Stot}) with
\begin{equation}
S_\Omega= -2\int \left( 
 Q \wedge  J _-
   + J_+
  \wedge M (Q ) \right)
  \label{SomJ}
\end{equation}
while  the 
alternative
coupling to the brane (\ref{sdfsr}) is 
\begin{equation}
\label{sdfggwg}
S'_\Omega= 2\int  
    F \wedge \ast J
\end{equation}
which agrees with (\ref{acth}) (up to a factor of 2 arising from the normalisation of the action).
In particular,
\begin{equation}
  F = d A + \ast J
\end{equation}
in agreement with the discussion in section \ref{Self}.

The transformations  (\ref{symom22}) with $\Omega =\ast J$   become
\begin{equation}
  \label{symom223} \delta P = \lambda, \quad  \delta J = \ast d  {\lambda} 
, \quad
  \delta Q = - \bar{\Pi}_+ d \lambda
\end{equation}
 and these leave the field strengths   invariant, $\delta F = \delta G = 0$.
 The variation of the action (\ref{Stot}) under these  follows from (\ref{delss2s}) and is
\begin{equation}
  \label{delss2ss} \delta S = -  \int \lambda \wedge d \ast J =-  \int \lambda \wedge  \ast j
\end{equation} 
 This is the same as the variation (\ref{erhdgh}) found in section \ref{Self} and vanishes provided
  the Dirac veto constraint (\ref{rhveto}) holds. Similarly, the variation of the alternative form of the interaction (\ref{sdfggwg}) has the same form.
 As a result, the theory has the expected generalised symmetries    as a result of  the Dirac veto.
 
 \vspace{.5cm}
 
 \noindent
{\bf {\large References}}

\end{document}